\documentclass[twocolumn,secnumarabic,amssymb, amsmath, nofootinbib,tightenlines,showpacs, aps, prl]{revtex4}

\usepackage{bm}%
\usepackage{graphics}%

\begin{document}
\input{epsf}
\def\Fig#1{Figure \ref{#1}}
\def\Eq#1{Eq.~(\ref{#1})}
\draft
\title{Oscillations in Rapid Fracture}
\author{Ariel Livne, Oded Ben-David and Jay Fineberg}
\address{The Racah Institute of Physics, The Hebrew
University of Jerusalem, Jerusalem 91904, Israel}
\begin{abstract}
Experiments of pure tensile fracture in brittle gels reveal a new
dynamic oscillatory instability whose onset occurs at a critical
velocity, $V_C=0.87C_S$, where $C_S$ is the shear wave speed. Until
$V_C$ crack dynamics are well described by linear elastic fracture
mechanics (LEFM). These extreme speeds are obtained by suppression
of the micro-branching instability, which occurs when sample
thicknesses are made comparable to the minimum micro-branch width.
The wavelength of these sinusoidal oscillations is independent of
the sample dimensions, thereby suggesting that a new intrinsic scale
exists that is unrelated to LEFM.
\end{abstract}
\pacs{46.50.+a, 62.20.Mk, 47.54.+r} \maketitle

The dynamics of rapidly moving cracks are of both fundamental and
practical importance. Driven by the elastic energy stored in a
stressed body, brittle cracks accelerate towards their theoretical
limiting speed \cite{Freund.90}. However in tensile fracture, the
crack tip becomes unstable at less than half that velocity (Rayleigh
wave speed - $V_R$), sprouting small side cracks termed
micro-branches \cite{Sharon PrB}. At this point, the single-crack
state is replaced by a multi-crack one and much of the energy
flowing to the crack tip, is transformed into additional fracture
surface (roughness). The density of micro-branches is observed to
increase with the main crack velocity and as a consequence, cracks
are rarely observed at velocities greater than $0.6-0.7 V_R$.

We describe experiments performed on polyacrylamide gels. Fracture
dynamics in these materials have been shown \cite{Livne} to be
identical to those of standard brittle amorphous materials, with the
advantage that the extremely slow wave speeds in gels simplify the
study of crack dynamics. We demonstrate that, when micro-branches
are suppressed, crack dynamics are well-described by the
single-crack equation of motion up to previously unattainable crack
speeds. Surprisingly, we find that a new oscillatory instability
occurs at a critical velocity close to $V_R$. The existence of this
instability highlights important unresolved questions regarding the
path selection of dynamic cracks \cite{Sethna, Path Prediction}.
Furthermore, the suppression of micro-branches when the gel
thickness approaches the process zone scale, raises interesting
questions regarding both the origin of the micro-branching
instability and the relevant length scale at which 3D fracture
becomes effectively 2D.

Our experiments were performed using sheets of brittle
polyacrylamide gels of typical size $(X \times Y)$ $ 125 \times
115mm^2$ and thickness $0.15 <d<0.35$mm where X, Y and Z are,
respectively, the propagation, loading, and sheet thickness
directions. All experiments reported here were performed in
polyacrylamide gels composed of 13.8\%  total monomer concentration
and 2.6\% bis-acrylamide as cross-linker \cite{Comment 2
concentrations}. These gels have a measured shear modulus of $G=35.2
\pm1.4 kPa$ yielding shear and longitudinal wave speeds of
$C_S=5.90\pm0.15m/s$ and $C_L=11.8\pm0.3m/s$, respectively.
Measurements of $C_S$ and $C_L$ using high-speed photography agree
with these values to within $0.1 m/s$. Experiments using other gel
compositions (e.g., 23\% total monomer concentration with 10\%
bis-acrylamide whose $C_S\sim 17m/s$) exhibit similar qualitative
and quantitative behavior. The gels were cast for each experiment
between flat glass plates. To ensure uniformity, the evaporation of
the gel's water content was limited to below 5\%. We compensated for
any resulting variations in wave speeds by direct measurement of $G$
for each experiment.

The experimental setup and measurement techniques are described
elsewhere \cite{Livne} and will be briefly reviewed here. The gel
sheets were loaded in Mode I (uniaxial extension) via constant
displacement in the vertical (Y) direction. The applied force was
monitored through a load cell. When a desired stress was reached, a
short cut was made at the sample's edge, midway between the vertical
boundaries. This seed crack either immediately or after a short
creep accelerated to dynamic velocities. Throughout each experiment
the crack tip's profile and immediate location were monitored by a
high speed camera set to a $X \times Y$ spatial resolution of $1280
\times 64/96$ pixels (frame rate of $7.2/5 kHz$). At the completion
of the fracture event, the fracture surface (XZ plane) was analyzed
via an optical microscope and the fracture profile (XY plane) was
scanned with 2400dpi resolution.

\begin{figure}
\centerline{\epsffile{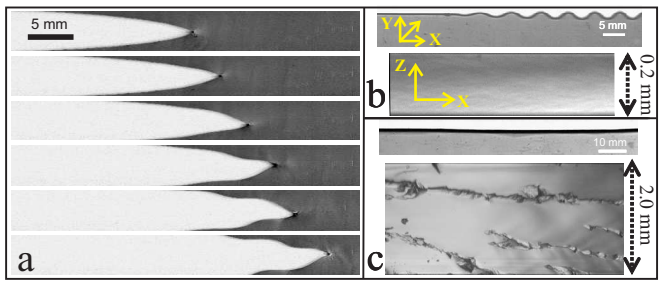}} \caption{(a) A sequence of
photographs, shown at  0.69 ms intervals, of a propagating crack
undergoing a transition from linear (top 2 pictures) to oscillatory
motion. Photographs of XY profile (top) and (XZ) fracture surface
(bottom) of (b) a 0.2 mm thick gel sample where oscillations
developed and  (c) a 2.0 mm thick gel where the crack retained its
straight line trajectory. In (c) the fracture surface is
micro-branch dominated whereas in (b) the oscillating crack has a
mirror surface. Propagation in (a), (b) and (c) was from left to
right.} \label{Ariel1}
\end{figure}

The dynamics of a typical experiment in thin gels are presented in
Fig. 1a. The initial crack propagates along a straight trajectory
until losing its directional stability to crack tip oscillations in
the XY plane. Straight-line propagation (Fig. 1b) typical of
micro-branch free fracture gives way to sinusoidal oscillations at a
millimeter scale. The fracture surface formed by oscillating cracks
is typically smooth (mirror-like), with rare instances of short
directed lines of successive micro-branches (``branch-lines'')
\cite{Sharon.02}. This behavior is in contrast to propagation in
thicker samples where the crack retains its straight line trajectory
throughout the fracture process (Fig. 1c). In these latter events,
the fracture surface is micro-branch dominated; beyond a critical
velocity branch-lines develop whose density increases with the crack
velocity \cite{Sharon.02}. Branch-lines disappear upon arrival at
the sample surfaces at Z=0 or Z=$d$. Thus, in thin sheets
branch-lines are quickly suppressed and sporadic.

The onset of the oscillations occurs at a critical velocity of
$V_C=0.87\pm0.02 C_S$ (Fig. 2a). The observed value of the critical
velocity is independent of $d$ for $0.15<d<0.3$mm. In thicker gels,
cracks rarely reached $V_C$ due to micro-branching (more on this
later). Surprisingly, no evidence of the Yoffe instability
\cite{Yoffe.51}, in which a crack is expected to deviate from its
original straight line trajectory at $V>0.6V_R$, is observed.

\begin{figure}
\centerline{\epsffile{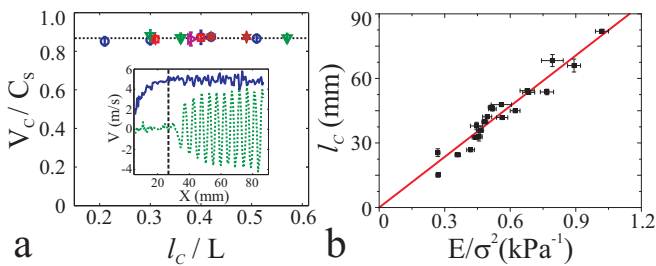}} \caption{(color online) (a)
Oscillations appear only beyond a critical velocity of
$V_C$=0.87$C_S$.  $V_C$ is independent of the crack length at the
onset, $l_c$, sample size, L, and thickness, $d$. The different
symbols represent measurements for $0.15<d< 0.30$ mm. (inset) The
velocity components, $V_x$ (continuous line) and $V_y$ (dotted
line), as a function of X, where the dashed line marks the onset of
oscillations. (b) The measured (squares) and predicted (line)
relation between $l_c$ and $E/\sigma^{2}$ in the framework of LEFM
\cite{Freund.90}. The line's slope gives the fracture energy
$\Gamma(v_c)=34\pm2 J/m^2$.}\label{Ariel2}
\end{figure}

The rapid acceleration to the extreme crack velocities attained
($V>0.87C_S$) suggests that waves reflected from the system's remote
boundaries can not affect crack dynamics, as reflected waves only
catch up with the crack tip well beyond the critical crack length
$l=l_c$. Thus, for $V \le V_C$, predictions by linear elastic
fracture mechanics (LEFM) of the equation of motion for a single
edge crack in a semi-infinite body \cite{Freund.90} should hold:
$V=V_R(1-\Gamma (v) E/K^2)$ where K=$1.12\sigma \sqrt{\pi l}$ and
$\Gamma$ is the material's fracture energy \cite{Lawn.93}. At
$V=V_C$ this predicts $l_c$ to be:
\begin{equation}
\l_c=\frac{\Gamma (v_c) E}{1.12^2 \pi \sigma ^2 (1 - V_C/V_R)}
\label{Eq1}
\end{equation}
Eq. \ref{Eq1} predicts that $l_c \propto \sigma ^{-2} $ where
$\sigma $ is the applied stress. This prediction (Fig. 2b) is,
indeed, in excellent agreement with measurements of $l_c$. The slope
of the linear fit yields a value of the fracture energy,
$\Gamma(V_C)=34\pm2 J/m^2$, which agrees with previous measurements
\cite{Fracture Energy of gels}. The slight scatter in Fig. 2b is
consistent with our 5\% allowed variation of $G$. This is the first
direct experimental verification of the equation of motion of a
single crack at velocities approaching $V_R$ \cite{Sharon Nature
1999}.

At $V_C$ the behavior of the crack's two velocity components, $V_x$
and $V_y$, changes abruptly. Prior to $V_C$,  $V_y = 0$ whereas
beyond $V_C$,  $V_y$  oscillates at a wavelength, $\lambda(x)$, with
increasing amplitude.  The oscillations in $V_y$ are accompanied by
a much smaller oscillatory component of $V_x$ having a wavelength of
$\lambda/2$ and a $\pi/2$ relative phase shift. (Fig. 2a inset).
This suggests that $V=\sqrt{V_x^2+V_y^2}$ does not, itself,
oscillate but increases monotonically, although we lack the temporal
resolution to directly verify this.

$V_x$ increases steadily until reaching $V_C$. Beyond $V_C$, the
full velocity, $V$, generally continues to accelerate (see e.g. Fig.
3) to steady state velocities. $V$, however, has not been observed
to surpass $C_S$ \cite{Comment 3 Max Speed}. Unexpectedly, the mean
value, over an oscillation period, of $V_x$ remains nearly {\em
constant} (even dropping by a few percent) upon reaching $V_C$. This
is illustrated in Fig. 3 (inset) where measured values of $V$ are
identical to values obtained by integrating the measured path length
of the crack and assuming that $V_x$ is constant, equal to its mean
value for $V>V_C$ \cite{Comment 4 Spatial Trajectory}.

\begin{figure}
\centerline{\epsffile{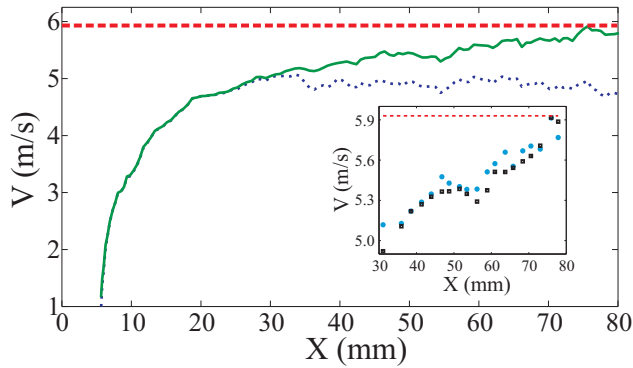}} \caption{(color online) The full
crack velocity, $V$ (continuous line), is compared to $V_x$ (dotted
line). Beyond $V_C$, $V$ increases until $C_S$ (dashed line)
\cite{Comment 3 Max Speed}, while $V_x$ stays nearly constant.
(inset) A comparison of direct measurements of $V$ beyond $V_C$
(full circles) with $V$, calculated using the spatial trajectory of
the oscillations (squares) \cite{Comment 4 Spatial Trajectory}. The
dashed line denotes $C_S$. All velocities are averaged over an
oscillation period.}\label{Ariel3}
\end{figure}

Let us consider the nature of the oscillations. The oscillations are
generally sinusoidal except for large amplitude states at the
highest values of applied stress, $\sigma$. Both the oscillation
amplitude, $A$, and wavelength, $\lambda$, develop with crack
advance until reaching steady-state values after a number of cycles
(Fig. 4 insets). The steady-state values of both $A$ and $\lambda$
vary with $\sigma$, as shown in Fig. 4, and do not appear to be
well-defined functions of $V$. At $V=V_C$, $A$ initiates from a
finite value, although no hysteresis in the transition to the
oscillatory state has been observed. $\lambda$ is a sharply
decreasing function of $\sigma$. Steady-state wavelengths range from
4-9mm with $V_x/\lambda \sim 0.5-1kHz$. This frequency scale
corresponds to the same ($\sim 1$ms) timescale that was previously
observed for micro-branching activation times in similar gels
\cite{Livne}. The origin of these length/frequency scales is
currently unknown. The time scale is over an order of magnitude
smaller than any possible resonance with the system's boundaries. In
fact, these scales are wholly {\em independent} of any of the
physical dimensions of the sample, since gels of substantially
different widths and dimensions all fall on the same steady-state
amplitude/wavelength curve (Fig. 4). Furthermore, as indicated by
Fig. 2b, the crack is effectively travelling in a semi-infinite
medium at the onset of oscillations and, under semi-infinite
conditions, the crack will have no knowledge of the actual size of
the sample. We, therefore, surmise that the length scale must have
its origins in an intrinsic scale, such as the process zone, that is
not quantitatively accounted for in current theoretical frameworks.

Although the steady-state values of $\lambda$ and $A$ are
well-defined, their initial transients  (Fig. 4 - insets) are
extremely diverse, with no discernable relation to either $V$ or
$\sigma$. It is interesting, however, that these transients follow a
number of distinct $A-\lambda$ trajectories, suggesting that there
may be a number of meta-stable solution branches that eventually
evolve to the steady-states presented in Fig. 4.

\begin{figure}
\centerline{\epsffile{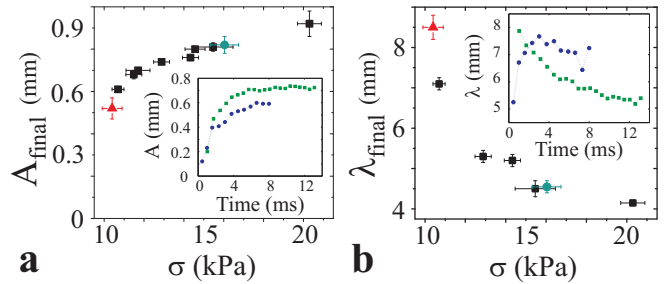}} \caption{(color online)
Steady-state amplitudes, $A$, (a) and wavelengths, $\lambda$, (b)
as a function of the applied stress, $\sigma$. (insets) Transient
time evolution of $A$ and $\lambda$ in two typical experiments.
t=0 defines the onset of oscillations. Sample dimensions: $(X
\times Y)$ $\sim 125 \times 115 mm^{2}$ ($\blacksquare$), $130
\times 155 mm^{2}$ ($\blacktriangle$) and $125 \times 70 mm^{2}$
($\bullet$). }\label{Ariel5}
\end{figure}

Why is the oscillatory instability not readily observed in thick
samples? Basically, in thick samples cracks rarely attain $V_C$, due
to the onset of micro-branching. Let us momentarily digress and
consider the relation of micro-branch formation to sample thickness.
Recent work in gels indicates that the micro-branching instability
is a first-order phase transition in which micro-branch formation is
triggered by noise, above a critical velocity \cite{Livne}.
Micro-branches in gels have a minimal width \cite{Livne}, $\Delta
Z$, whose scale (here $\sim 40\mu m$) is approximately that of the
process zone. The system's quenched noise level may then be related
to $d/\Delta Z$. This may explain why the number of branching events
is strongly suppressed with decreasing $d$. This suppression,
coupled with the disappearance of branch-lines with their arrival at
the sample faces at $Z=0$ or $d$ (see e.g. Fig. 1c), causes the
increasingly sporadic appearance of micro-branches as $d \rightarrow
\Delta Z$. Thus for $d \rightarrow \Delta Z$ ($\Delta Z \sim \mu m$
for PMMA and less for glass) micro-branching events are rare with
crack dynamics becoming effectively two-dimensional. This
suppression of micro-branches explains how oscillations are so
easily obtained for $d \sim 0.15-0.2mm$.

Although rare, the micro-branching instability {\em is} observed in
thin sheets both in straight and oscillatory propagation. When
excited, micro-branches reduce the energy available to the main
crack \cite{Sharon PrB}, and consequently reduce the main crack's
velocity. If $V$ falls below $V_C$, oscillations immediately cease,
and the crack propagates along a straight trajectory (e.g. Fig. 5 in
the 2nd shaded region). When micro-branching is sporadic,
micro-branches only momentarily reduce $V$ to below $V_C$. This not
only causes discontinuities in the oscillation phase but before and
after a velocity drop we often observe entirely different sinusoidal
solutions (with significantly different wavelength and amplitude).
When {\em prolonged} micro-branching occurs, as characteristic of
the thicker gels, $V$ only momentarily surpasses $V_C$ between
micro-branching events. In this case, pronounced ``stair-case"
profiles are created by momentary diversions of the crack for less
than a period followed by straight line trajectories. All of these
effects are demonstrated in the single experiment, presented in Fig.
5, in which $d$ is tapered as a function of $X$. The rarity of
micro-branches in the initial stage ($d < 0.35$mm) enables rapid
acceleration to $V>V_C$ and the onset of oscillations. As the taper
widens, micro-branching occurs; first sporadically (2nd shaded
region) and later with prolonged branching (for $d>0.35$mm), leading
to stair-case profiles.

\begin{figure}
\centerline{\epsffile{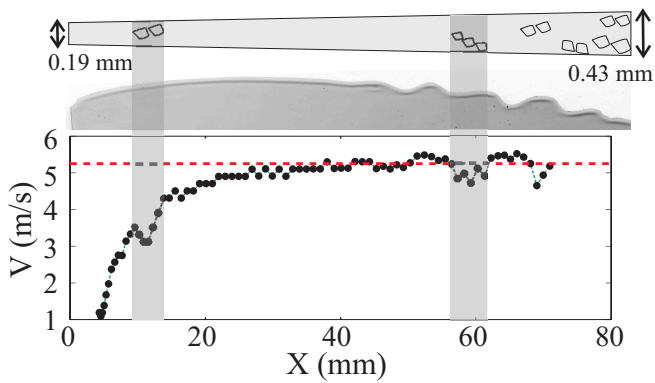}} \caption{(color online) Tapering of
$d$ ($0.19<d<0.43$mm) as a function of $X$ enables $V>V_C$ for
thicknesses ($d>0.35mm$) normally dominated by micro-branches. (top)
Schematic drawing of the XZ fracture surface indicating branch-line
locations and $d(x)$. (center) Photograph of the crack profile along
the XY plane with (bottom) the corresponding velocity profile.
Shaded rectangles indicate regions of sporadic branch-lines (diamond
shapes at top) and the resulting drops in $V$ until their
disappearance when the branch-line encounters the $Z=0$ or $d$
planes. When $V<V_C$ ($V_C$ denoted by the dashed line) oscillations
cease with no hysteresis and cracks continue along a straight-line
trajectory (e.g. second shaded region). Prolonged branching
($d>0.35$mm) yields ``staircase" profiles. }\label{Ariel4}
\end{figure}

In recent years, two oscillatory instabilities of cracks traveling
in thin sheets have been reported. Both of these are fundamentally
different from the oscillatory instability presented here. Deegan
and coworkers have observed wavy cracks traveling at intersonic
velocities in biaxially-stretched thin rubber sheets \cite{Cracks in
Rubber}. In our experiments, oscillations are observed at clearly
{\em subsonic} velocities in pure uniaxial tension. Phase field
models \cite{Phase field}, which do yield subsonic oscillations of
crack trajectories, predict $\lambda$ to scale with the system size,
also in sharp contrast with our results. Recent calculations
\cite{Procaccia}, pairing LEFM with the Hodgdon and Sethna
\cite{Sethna} path selection criterion, predict the onset of
oscillatory instability at close to our measured values of $V_C$. In
contrast to our measurements however, $\lambda$ is predicted to
scale with the system size.

Oscillatory cracks have also been observed when a rigid cutting tool
is forced through a thin elastic sheet \cite{Oscillations by driving
a rod}. These cracks travel at $V << C_S$ with out-of-plane bending
playing an important role. In contrast, the instability presented
here is in pure Mode I with no measurable out-of-plane deflection
\cite{Comment 6 Mode III}. Our oscillations are triggered at $V_C$
values that are near $C_S$, irrespective of the sample's thickness.
In addition, the crack dynamics in our experiments are consistent
with the single-crack (Mode I) equation of motion predicted by LEFM
until either $V_C$ or the onset of micro-branching. We therefore
conclude that these observations indicate a previously unobserved
instability in dynamic fracture whose characteristic time/length
suggests that a new {\em intrinsic} scale is important to describe
these dynamics. This time/length scale can not be explained in the
framework of LEFM \cite{Procaccia}, underlining the necessity for a
more fundamental theoretical description of the near vicinity of the
crack tip.

This research was supported by the Israel Science Foundation (grant
194/02).


\begin{thebibliography}{10}
\bibitem{Comment 3 Max Speed}
The limiting velocity should be $V_R= 0.955C_S$ and not $C_S$. In a
number of experiments, the crack's maximal velocity was between
$V_R$ and $C_S$. This discrepancy might be due to measurement
accuracy ($\sim$3\%) or a different form for $V_R$ in elastomers/
thin sheets.
\bibitem{Comment 6 Mode III}
The relevance of mode III fracture to these oscillations was ruled
out in the following manner: With a laser sheet we illuminated a
thin line on the gel which was oriented parallel to the Y axis at a
point midway along the X axis, where (future) crack oscillations
appear. If the crack tip opening in the vicinity of this line were
in mode III, we would expect to observe deflection of the line, when
traversed by the crack. No such deflection was observed to the $<
5^\circ$ measurement accuracy.
\end{thebibliography}
\end{document}